\documentclass[3p,preprint]{elsarticle}

\usepackage{color}
\usepackage[utf8]{inputenc}
\usepackage{graphicx}
\usepackage{amsmath}
\usepackage{amssymb}
\usepackage{longtable}
\usepackage{footnote}
\makesavenoteenv{tabular}
\makesavenoteenv{table}
\usepackage{threeparttable}

\usepackage{hyperref}
\hypersetup{
  pdfnewwindow=true, colorlinks=true,
  linkcolor=blue, anchorcolor=blue,
  citecolor=blue, filecolor=blue,
  menucolor=blue, urlcolor=blue}

\DeclareMathOperator{\Ima}{Im}
\DeclareMathOperator{\Tr}{Tr}

\graphicspath{{./Fig/}}

\newcommand{\comment}[1]{\textit{}}
\newcommand{\bit}{\begin{itemize} \setlength{\itemsep}{0ex} \setlength{\topsep}{0ex} } 
\newcommand{\eit}{\end{itemize}}
\newcommand{\be}{\begin{equation}}
\newcommand{\ee}{\end{equation}}
\newcommand{\bea}{\begin{eqnarray}}
\newcommand{\eea}{\end{eqnarray}}
\newcommand{\ba}{\begin{align}}
\newcommand{\ea}{\end{align}}
\newcommand{\SKIP}[1]{}

\DeclareMathAlphabet\mathbfcal{OMS}{cmsy}{b}{n}

\begin{document}

\begin{frontmatter}

\author[ETH]{QuanSheng Wu\corref{cor1}}\ead{wuq@phys.ethz.ch}
\cortext[cor1]{Corresponding author}
\author[IOP]{ShengNan Zhang}
\author[CAEP]{Hai-Feng Song}
\author[ETH]{Matthias Troyer}
\author[ETH,StP]{Alexey A. Soluyanov}
\address[ETH]{Theoretische Physik and Station Q Zurich, ETH Zurich, 8093 Zurich, Switzerland}
\address[IOP]{Institute of Physics, Chinese Academy of Sciences, Beijing 100190}
\address[CAEP]{Institute of Applied Physics and Computational Mathematics, Beijing 100094, China}
\address[StP]{Department of Physics, St. Petersburg State University, St. Petersburg, 199034 Russia}

\date{\today}
\title{\texttt{WannierTools}: An open-source software package for novel topological materials}

\begin{abstract}
We present an open-source software package \texttt{WannierTools}, a tool for investigation of novel topological materials. This code works in the tight-binding framework, which can be generated by another software package Wannier90~\cite{Wannier90cpc}. It can help  to classify the topological phase of a given materials by calculating the Wilson loop, and can get the surface state spectrum which is detected by angle resolved photoemission (ARPES)  and in scanning tunneling microscopy (STM) experiments . It also identifies positions of Weyl/Dirac points and nodal line structures,  calculates the Berry phase around a closed momentum loop and Berry curvature in a part of the Brillouin zone(BZ).  
\end{abstract}


\end{frontmatter}

\noindent\textbf{PROGRAM SUMMARY}

\noindent\textit{Program title:} \texttt{WannierTools}

\noindent\textit{Catalogue identifier:} TO BE DONE

\noindent\textit{Program summary URL:} TO BE DONE

\noindent\textit{Program obtainable from:} CPC Program Library, Queen’s University, Belfast, N. Ireland

\noindent\textit{Licensing provisions:} GNU General Public Licence 3.0

\noindent\textit{No. of lines in distributed program, including test data, etc.:} 17118 lines

\noindent\textit{No. of bytes in distributed program, including test data, etc.:} 7421589 bytes

\noindent\textit{Distribution format:} tar.gz

\noindent\textit{Programming language:} Fortran 90

\noindent\textit{Computer:} any architecture with a Fortran 90 compiler

\noindent\textit{Operating system:} Unix, Linux, Mac OS X

\noindent\textit{Has the code been vectorised or parallelized?:} Yes

\noindent\textit{RAM:} Variable, depends on the complexity of the problem

\noindent\textit{Classification:} 40

\noindent\textit{External routines/libraries used:} 
\begin{itemize}
\item BLAS (http://www/netlib.org/blas)
\item LAPACK (http://www.netlib.org/lapack)
\end{itemize}

\noindent\textit{Nature of problem:} Identifying the topological classifications of crystalline systems including insulators, semimetals, metals, and studying the electronic properties of the related slabs and ribbons systems. 

\noindent\textit{Solution method:}  Tight-binding method is a good approximation for solid system. Based on that, Wilson loop is used for topological phase classification. The iterative Green's function is used for obtaining the surface state spectrum. 

\noindent\textit{Running time:} Depends on the complexity of the problem

\section{Introduction}
Novel topological states have attracted much attention during the past decades. They give a lot of opportunities to explore new physics and to realize new quantum devices~\cite{RevModPhys.83.1057,RevModPhys.82.3045}. Since quantum spin hall effect (QSHE) was predicted in 2005~\cite{PhysRevLett.95.226801} , and  realized in HgTe/CdTe quantum wells in 2007~\cite{Bernevig1757, Konig766}, more and more topological novel states were discovered, such as 3D topological insulators~\cite{Hsieh2008, Zhang2009, Xia2009}, Dirac~\cite{PhysRevB.85.195320, Wang2013}, Weyl~\cite{PhysRevB.83.205101, PhysRevX.5.011029, Soluyanov2015} semimetals, Hourglass fermions~\cite{Wang2016}, Nodal line semimetals ~\cite{PhysRevB.84.235126, PhysRevLett.115.036807, PhysRevB.93.121113}  and nodal chain metals~\cite{Bzdusek2016} et al... Nowadays, a lot of new topological phases are emerging, and more and more materials are identified to be topologically non-trivial. 

One important feature of topological materials is topologically protected surface~\cite{surface} states, which are robust against the white noise disorder and have novel transport properties~\cite{Konig766} due to spin-momentum locked features. In experiments, the surface states and transport properties are detectable, therefore, taken as an evidence for non-trivial topology in bulk band structures ~\cite{Konig766, Hsieh2008, Hasan2010}. The logic here is that the bulk-edge correspondence principle, which  tells that there are topologically protected $d-1$-dimension edge states if the topological property is non-trivial in  $d$-dimension bulk. In the theoretical part, besides calculation of surface-state spectrum and transport properties, we could also calculate  topological numbers such as the $\mathbb{Z}_2$ number~\cite{Kane2015_z2}, Chern number~\cite{TKNN1982}  and Wilson loops~\cite{Yu2011_z2} (identical to Wannier charge centers~\cite{Soluyanov2011_z2}) to study the topology of energy band structures of a material directly.  

At present, there are more and more groups joining the investigation of novel topological materials. However, there are only a few software packages that can be used so far. Z2pack~\cite{z2pack} is a package that uses the Wannier charge centers (WCCs) to classify the topological properties of real compounds. By using Z2pack, the WCCs can be obtained either from tight-binding (TB) model or from  first-principle packages, such as VASP, ABINIT and Quantum-espresso. PythTB ~\cite{pythtb} is another software package with a Python implementation of TB models. There are many functions to build and solve TB models of the electronic structure of systems for arbitrary dimensional systems (crystals, slabs, ribbons, clusters, etc.) in PythTB, and it can compute Berry phases and related properties. 

Here we introduce a TB dependent open-source software package called \texttt{WannierTools}, which can be used for novel topological materials  investigation. Unlike Z2pack, it can be used to calculate the surface states of materials, and, being parallelized with MPI, it is faster than PythTB. It is  a user friendly and efficient single program implemented in Fortran90. The only thing needed to run it is an input file, which contains some parameters describing  your systems, and a TB model written in Wannier90\_hr.dat format~\cite{Wannier90cpc}.  With \texttt{WannierTools}, topological numbers like the $\mathbb{Z}_2$ numbers or Wilson loops for the bulk system can be calculated in order to explore the topological properties of a material.  It can also help  to search  for Weyl/Dirac points or nodal loop structures in the BZ of metallic systems. There are plenty of other functions, e.g. studying the electron structure properties for slab and ribbon systems, studying the Berry curvature for bulk systems, studying the Berry phase around one momentum loop in the BZ for nodal-line systems and so on.

This paper is organized as follows. In section~\ref{sec:method}, we review briefly some basic theories related to this package. In section~\ref{sec:functions}, we introduce the capabilities of this package. In section~\ref{sec:install}, we introduce the installation and basic usages.  In section~\ref{sec:examples}, we introduce  a new topological  material HfPtGe in order to show you how to use \texttt{WannierTools} to explore a new topological phase.


\section{Methods} \label{sec:method}
\subsection{TB method}\label{subsec:tb}
TB method is a semi-empirical approach to study   electronic structures of solid-state systems by projecting the Hamiltonian of the system onto a series of  local orbitals. There are several ways to construct TB models, such as Slater-Koster method~\cite{PhysRev.94.1498}, maximum localized Wannier functions(MLWF)~\cite{Marzari_RMP}, and  discretization of  $k\cdot p$ model~\cite{Willatzen2009} onto a lattice. Among these methods, the MLWF method\cite{Marzari_RMP} is widely used by the people who are interested in real materials simulations. MLWF is implemented in Wannier90, which has many interfaces with different first-principle software packages like VASP, WIEN2k, et al.. Therefore MLWF TB models can be automatically obtained from  first-principle calculations together with Wannier90~\cite{Wannier90cpc}. 

In different TB methods, the basis functions could be mutually orthogonal or non-orthogonal. However,  \texttt{WannierTools} is only capable of dealing with the TB models with orthogonal basis functions. Fortunately, the Wannier functions (WFs) for MLWF TB method fulfill this limitation. In this section, we give some brief introductions to the general orthogonal TB methods. The details of how to construct MLWF TB models can be found in Refs.~\cite{Mostofi20142309,Marzari_RMP}

Let $i$ label the atoms, $\mu$ label the orbitals, $m$ label the combination of  \{$i\mu$\}, $\boldsymbol R$ label the lattice vectors in 3D crystal, and $\boldsymbol \tau_i$ label the position of atoms in a home unit cell.  The local orbital for the i'th atom centered at $\boldsymbol R+\boldsymbol\tau_i$ can be written as 
\begin{eqnarray}
\phi_{\boldsymbol R m}( \boldsymbol r) \equiv \phi_m(\boldsymbol r- \boldsymbol R) \equiv \varphi_{i\mu}(\boldsymbol r- \boldsymbol R- \boldsymbol \tau_i)
\end{eqnarray}
The orthogonality of orbitals requires $\langle \phi_{\boldsymbol R m} |\phi_{\boldsymbol R' n}\rangle=\delta_{{\boldsymbol {RR'} }}\delta_{mn}$.  TB parameters of the Hamiltonian that have the translational symmetry due to  Bloch theorem can be calculated via
\begin{eqnarray}
H_{mn}(\boldsymbol R) = \langle \phi_{\boldsymbol 0 m}|\hat{H}|\phi_{\boldsymbol R n}\rangle 
\end{eqnarray}
Once we have the TB Hamiltonian $H_{mn}(\boldsymbol R) $, the Hamiltonian in  k space can be obtained by a Fourier transformation (FT) ~\cite{Marzari_RMP}. There are two conventions~\cite{pythtb} for FTs.  One is 
\begin{eqnarray}
H_{mn}(\boldsymbol k)= \sum_{\boldsymbol R} e^{i\boldsymbol k \cdot \boldsymbol R} H_{mn}(\boldsymbol R) \label{eqn:Hconv1}
\end{eqnarray} 
The other one is 
\begin{eqnarray}
H_{mn}(\boldsymbol k)= \sum_{\boldsymbol R} e^{i\boldsymbol k \cdot (\boldsymbol R+\boldsymbol \tau_m- \boldsymbol\tau_n)} H_{mn}(\boldsymbol R)\label{eqn:Hconv2}
\end{eqnarray} 
It can be demonstrated that eigenvalues for these two conventions are the same, but the eigenvectors are different. The eigenvectors of the first convention Eqn.(\ref{eqn:Hconv1}) are analogous to the Bloch wave functions $\psi_{n\boldsymbol k}(\boldsymbol r)$. The eigenvectors of the second convention  Eqn.(\ref{eqn:Hconv2}) are analogous to the periodic part of the Bloch wave functions $u_{n\boldsymbol k}(\boldsymbol r)=\psi_{n\boldsymbol k}(\boldsymbol r)e^{-i\boldsymbol k \boldsymbol r}$, which is of great importance in Berry phase and Berry curvature or the Wannier centers calculations. Therefore, the second convention is used in \texttt{WannierTools}.

According to the bulk-edge correspondence, there are topologically protected surface states if the topology of bulk energy bands is non-trivial. In order to study such surface states, we have to construct a slab system which is periodic along two directions at the surface. In practice, a new unit cell is defined with lattice vectors $\bold{R'}_{1,2,3}$, 
\begin{eqnarray}
\bold{R'}_1= U_{11}\,\bold{R}_1+U_{12}\,\bold{R}_2+U_{13}\,\bold{R}_3\\
\bold{R'}_2= U_{21}\,\bold{R}_1+U_{22}\,\bold{R}_2+U_{23}\,\bold{R}_3\\
\bold{R'}_3= U_{31}\,\bold{R}_1+U_{32}\,\bold{R}_2+U_{33}\,\bold{R}_3
\end{eqnarray}
Where $\bold{R}_{1,2,3}$ are lattice vectors of the original unit cell of the bulk system, $\bold{R'}_1$ and $\bold{R'}_2$ are two lattice vectors in the target slab surface,  $\bold{R'}_3$ is the other lattice vector which is out of the surface and fulfills the volume fixed condition,
\begin{eqnarray}
 \bold{R'}_1 \cdot ( \bold{R'}_2  \times \bold{R'}_3)= \bold{R}_1 \cdot ( \bold{R}_2  \times \bold{R}_3)
\end{eqnarray}
Since the slab system is non-periodic along the $\bold{R'}_3$ direction, the Hamiltonian of a slab system with a 2D momentum $k_{\parallel}$ can be calculated by the following FT
\begin{eqnarray}
H^{slab}_{mn}(\bold{k}_{\parallel} )= \bold{\sum_{||\bold{R}||}} e^{i\, \bold{k}_{\parallel}\cdot \bold{R}}  H^{slab}_{mn}(\bold{R})
\end{eqnarray}
Where $\bold{R}=a'\bold{R'}_3+b'\bold{R'}_2+c'\bold{R'}_3$ and $||\bold{R}||$ is a restriction that the summation is only carried on  $a'$ and $b'$ with different $c'$. We label the layer index along $\bold{R'}_3$ as $i,j$. As a consequence, the Hamiltonian of a slab system with $n_s$ layers can be written in the layer index matrix form,
\begin{eqnarray}
H^{slab}_{mn}(\bold{k}_{\parallel} )= 
\left(\begin{array}{cccc}
H_{mn,11}(\bold{k}_{\parallel} ) & H_{mn,12}(\bold{k}_{\parallel} ) & \cdots & H_{mn,1n_s}(\bold{k}_{\parallel} )  \\
H_{mn,21}(\bold{k}_{\parallel} )& H_{mn,22}(\bold{k}_{\parallel} ) & \cdots & H_{mn,2n_s}(\bold{k}_{\parallel} )  \\
\vdots&\vdots  & \ddots &  \vdots \\
H_{mn,n_s1}(\bold{k}_{\parallel} ) & H_{mn,n_s2}(\bold{k}_{\parallel} ) & \cdots& H_{mn,n_sn_s}(\bold{k}_{\parallel} )
\end{array}\right) \label{eqn:hslab}
\end{eqnarray}
where the diagonal elements of the Hamiltonian are the intra-plane ones, and the off diagonal elements are the inter-plane ones. The element in  Eq. (\ref{eqn:hslab})  can be read explicitly as, 
\begin{eqnarray}
 H_{mn,ij}(\bold{k}_{\parallel})=\sum_{\bold{R}=\left\{\bold{R'}_1, \bold{R'}_2, (i-j)\bold{R'}_3 \right\} }
 e^{i\, \bold{k}_{\parallel}\cdot \bold{R}}  H_{mn}(\bold{R})
\end{eqnarray}
Finally the energy band of the slab system can be obtained straightforwardly to diagonalize the Eq. (\ref{eqn:hslab}). 

By the way, the Hamiltonian for the ribbon system can be obtained in the same way as the slab system does. The difference is that there are two confined directions $\bold{R'}_1, \bold{R'}_2$ in ribbon systems, which enlarges the size of Hamiltonian.

\subsection{Wannier charge center calculation}
$\mathbb{Z}_2$  topological number~\cite{Kane2015_z2} and Chern number~\cite{TKNN1982} are applied to classify  topological properties for time-reversal invariant and time-reversal symmetry breaking systems respectively. In inversion symmetric invariant system, the $\mathbb{Z}_2$  topological number can be calculated by multiplying the parities of the occupied bands at  time reversal invariant momenta (TRIMs) in the Brillouin zone~\cite{Fu2007}.    There are several methods~\cite{Takahiro2007_Z2, Yu2011_z2, Soluyanov2011_z2} to calculate the $\mathbb{Z}_2$  number in inversion symmetry breaking systems. Among them it was demonstrated that the  Wilson loop~\cite{Yu2011_z2} and Wannier charge centers (WCCs)~\cite{Soluyanov2011_z2} method are equivalent to each other, and  are also valid  for time-reversal symmetry breaking systems. In \texttt{WannierTools}, we take the algorithm presented in Refs.~\cite{Soluyanov2011_z2,z2pack}. The hybrid WFs~\cite{PhysRevLett.102.107603} are defined as
\begin{eqnarray}
|nk_xl_y\rangle =\frac{1}{2\pi} \int_0^{2\pi}d k_y e^{-ik_yl_y}|\psi_{n{\bold k}}\rangle
\end{eqnarray}
where $|\psi_{n{\bold k}}\rangle$ is the Bloch wave function. The hybrid Wannier centers are defined as
\begin{eqnarray}
\bar{y}_n(k_x)&=& \langle nk_x0|y|n k_x0\rangle \\
  &=&\frac{i}{2\pi} \int _{-\pi} ^{\pi}d {k_y} \langle u_{n,k_x, k_y}|\partial _{k_y}| u_{n,k_x, k_y}\rangle \label{eqn:wcc}
\end{eqnarray}
where $| u_{n,k_x, k_y}\rangle$ is the periodic part of Bloch function $|\psi_{n{\bold k}}\rangle$. In practice , the integration over $k_y$ is transformed  by a summation over the discretized $k_y$. Eqn.(\ref{eqn:wcc}) can be reformulated using the discretized Berry phase formula~\cite{Marzari1997}, 
\begin{eqnarray}
\bar{y}_n(k_x)&=& -\frac{1}{2\pi} \text{Im} \ln \prod_j M_{nn}^{(j)} \label{eqn:wcc2}
\end{eqnarray}
where the gauge-dependent overlap matrix $M_{mn}^{(j)}=\langle u_{m,k_x, k_{y_j}}| u_{n,k_x, k_{y_{j+1}}} \rangle$ is introduced. However, the summation of the hybrid Wannier centers $\bar{y}_n(k_x)$ over $k_x$ is gauge invariant~\cite{PhysRevB.47.1651}, which is the total electronic polarization. As shown in Ref~\cite{Marzari1997, Soluyanov2010}, there is another way to obtain $\bar{y}_n(k_x)$. Firstly, we get the "unitary part" $\tilde{M}^{(j)}$ of each overlap matrix $M_{mn}^{(j)}$ by carrying out the single value decomposition $M=V\Sigma W^{\dag}$, where V and W are unitary and $\Sigma$ is real-positive and diagonal. Then we  set $\tilde{M}^{(j)}=VW^{\dag}$. The eigenvalues $\lambda_n$ of matrix $\Lambda=\prod_j\tilde{M}^{(j)}$ are all of unit modulus. The hybrid Wannier centers are defined with the phases of $\lambda_n$ eventually,  
\begin{eqnarray}
\bar{y}_n(k_x)=-\frac{1}{2\pi} \text{Im} \ln \lambda_n 
\end{eqnarray}
We can get the topological properties of $k_x-k_y$ plane from the evolution of $\bar{y}_n(k_x)$ along a $k_x$ string. The detail of such classification of WCCs or Wilson loop are discussed in Refs~\cite{Yu2011_z2,Soluyanov2010,Soluyanov2011_z2}. More information can be found in Ref ~\cite{z2pack}. 

\subsection{Berry phase and Berry curvature}
In this section, we give the basic formalism for computing Berry phase~\cite{PhysRevLett.62.2747,RevModPhys.82.1959} and Berry curvature~\cite{PhysRevLett.49.405,0953-8984-20-19-193202} of Bloch states. Firstly, we introduce the single band case, where the energy bands are isolated to each other. A Berry phase $\phi_n$ is a geometric phase associated with the phase evolution of the n'th state over a closed curve $\mathcal{C}$ in external parameter space k, defined as
\begin{eqnarray}
\phi_n=\oint_{\mathcal{C}} {{\boldsymbol A}_n}\cdot d{ \boldsymbol k} 
\end{eqnarray}
Where the Berry connection is $A_{n, \alpha}= i \langle u_{n, { \boldsymbol k}} |\partial _{\alpha} u_{n, { \boldsymbol k}}\rangle, \alpha=k_x, k_y, k_z$, and Berry curvature are introduced
\begin{eqnarray}
\Omega_{n, \alpha\beta}&=& \partial_{\alpha}A_{n, \beta}- \partial_{\beta}A_{n, \alpha} \\
&=& i\langle\partial_{\alpha} u_{n, { \boldsymbol k}} |\partial _{\beta} u_{n, { \boldsymbol k}}\rangle
-i\langle\partial_{\beta} u_{n, { \boldsymbol k}} |\partial _{\alpha} u_{n, { \boldsymbol k}}\rangle\\
&=&-2 \Ima \langle\partial_{\alpha} u_{n, { \boldsymbol k}} |\partial _{\beta} u_{n, { \boldsymbol k}}\rangle
\end{eqnarray} 
Secondly, for multi-band case, it is often to treat the occupied $N_{occ}$ bands as a joint band manifold, which is referred to as the "non-Abelian" case. Generalizations for the formalism of  Berry phase and Berry curvature from single band to multi-band case are as follows, 
\begin{eqnarray}
\phi= \oint_{\mathcal{C}} \Tr[{{ \mathbfcal{A}}}]\cdot d{ \boldsymbol k} \label{eqn:berryphase}
\end{eqnarray}
Where the Berry connection ~\cite{RevModPhys.82.1959} is $\mathcal{A}_{mn, \alpha}= i \langle u_{m, { \boldsymbol k}} |\partial _{\alpha} u_{n, { \boldsymbol k}}\rangle, \alpha=k_x, k_y, k_z$, and Berry curvature is
\begin{eqnarray}
\Omega_{mn, \alpha\beta}&=& \partial_{\alpha}\mathcal{A}_{mn, \beta}- \partial_{\beta}\mathcal{A}_{mn, \alpha} -i[\mathcal{A}_{\alpha}, \mathcal{A}_{\beta}]_{mn} \\
&=& i\langle\partial_{\alpha} u_{m, { \boldsymbol k}} |\partial _{\beta} u_{n, { \boldsymbol k}}\rangle
-i\langle\partial_{\beta} u_{m, { \boldsymbol k}} |\partial _{\alpha} u_{n, { \boldsymbol k}}\rangle\\
&=&-2 \Ima \langle\partial_{\alpha} u_{m, { \boldsymbol k}} |\partial _{\beta} u_{n, { \boldsymbol k}}\rangle
\end{eqnarray}
and define
\begin{eqnarray}
\Omega_{\alpha\beta}=\Tr\Omega_{mn, \alpha\beta}
\end{eqnarray}
where $\Tr$ denotes a trace over the occupied bands. 

In practice, the integration of Eqn.(\ref{eqn:berryphase}) is implemented on a discrete k mesh. The loop $\mathcal{C}$ is discretized into a series of closely space points $k_j$. Accordingly, the Berry phase becomes
\begin{eqnarray}
\phi= -\sum_{j}\Ima \ln \det M^{(j)}= -\Ima \ln \prod_j \det M^{(j)}
\end{eqnarray}
where the overlap matrix $M_{mn}^{(j)}$ is the same as in Eqn.(\ref{eqn:wcc2}), i.e., $M_{mn}^{(j)}=\langle u_{m,{ \boldsymbol k}_j}| u_{n,{ \boldsymbol k}_{j+1}} \rangle$.

\subsection{Calculation of surface states}
Theoretically, we have two methods to get surface spectrum corresponding to the bulk topology. One is that we calculate the band structure of a slab system, which was introduced in section \ref{subsec:tb}. The other one is to calculate the surface Green's function (SGF) for a semi-infinite system that will be introduced in this section. In the 1970s, one of the most popular GF approachs  was based on the "effective field" and transfer matrix~\cite{Falicov1975, PhysRevB.23.4988, PhysRevB.23.4997}, which are relatively of slow convergence especially near singularities. Now, the extensively used schemes to obtain the SGFs is the iterative Green's function method developed in the 1980s~\cite{PhysRevB.28.4397, Sancho1984}.  With an effective concept of principle layers (The layer that is large enough to guarantee that hoppings between the next nearest layers are negligible.), the iterative procedure can save quite an amount of computational time. The method~\cite{PhysRevB.28.4397, Sancho1984} involves replacing the principle layer by an effective two principle layers,  and these effective layers interact through energy-dependent residual interactions which are weaker than those of the original ones. This replacement can be repeated iteratively until the residual interactions between the effective layers become as small as desired. Each new iteration doubles the number of the original layers included in the new effective layer. That is, after $n$ iterations, one has a chain of lattice constant $2^n$ times the original one, and each effective layer replacing $2^n$ original layers. The details of the algorithm are presented in Ref~\cite{Sancho1985}. For the integrity of the paper, we list the main iterations that we reused in \texttt{WannierTools}. The most important parameters for iteration $i$ are the following
\begin{eqnarray}
&&\alpha_i= \alpha_{i-1}(\omega-\varepsilon_{i-1})^{-1} \alpha_{i-1}\nonumber\\
&&\beta_i= \beta_{i-1}(\omega-\varepsilon_{i-1})^{-1} \beta_{i-1}\nonumber\\
&&\varepsilon_i=\varepsilon_{i-1}+\alpha_{i-1}(\omega-\varepsilon_{i-1})^{-1} \beta_{i-1}+\beta_{i-1}(\omega-\varepsilon_{i-1})^{-1} \alpha_{i-1}\label{eqn:itergreen}\\
&&\varepsilon_i^s=\varepsilon_{i-1}^s+\alpha_{i-1}(\omega-\varepsilon_{i-1})^{-1} \beta_{i-1}\nonumber\\
&&\tilde{\varepsilon}_i^s=\tilde{\varepsilon}_{i-1}^s+\beta_{i-1}(\omega-\varepsilon_{i-1})^{-1} \alpha_{i-1}\nonumber
\end{eqnarray}
with the initialization $\varepsilon_0=\tilde{\varepsilon}_0=\tilde{\varepsilon}_0^s=H_{00}({\bold k}_{\parallel})$, $\alpha_0=H_{01}({\bold k}_{\parallel})$, $\beta_0=H_{01}^{\dag}({\bold k}_{\parallel})$, where $H_{00}({\bold k}_{\parallel})$ is the intra-hopping parameters in the principle layers, $H_{01}({\bold k}_{\parallel})$ is the inter-hopping parameters between the nearest neighbor of principle layers. 

Iteration of Eqn.(\ref{eqn:itergreen}) should be converged until  ${\varepsilon}_n \simeq {\varepsilon}_{n-1}$  and $\tilde{\varepsilon}_n^s \simeq \tilde{\varepsilon}_{n-1}^s$. The SGFs $G_{s}({\bold k}_{\parallel}, \omega)$ and the bulk GF $G_{b}({\bold k}_{\parallel}, \omega)$  can be obtained as
\begin{eqnarray}
&&G_{s}({\bold k}_{\parallel}, \omega)\simeq (\omega-\varepsilon_n^s)^{-1} \\
&&\tilde{G}_{s}({\bold k}_{\parallel}, \omega)\simeq (\omega-\tilde{\varepsilon}_n^s)^{-1} \\
&&{G}_{b}({\bold k}_{\parallel}, \omega)\simeq (\omega-\varepsilon_n)^{-1} 
\end{eqnarray}
Where $\tilde{G}_{s}$ is the SGF of the dual surface. The surface spectrum function $A({\bold k}_{\parallel}, \omega)$ can be obtained from the imaginary part of SGF
\begin{eqnarray}
A({\bold k}_{\parallel}, \omega)= -\frac{1}{\pi} \lim_{\eta \to 0^+}\Ima \Tr G_s({\bold k}_{\parallel}, \omega+i \eta)
\end{eqnarray}
The spin texture of surface states  can also be obtained with~\cite{Dai2007} 
\begin{eqnarray}
{\bold S}({\bold k}_{\parallel}, \omega)= -\frac{1}{\pi} \lim_{\eta \to 0^+}\Ima \Tr \left[ {\boldsymbol \sigma} G_s({\bold k}_{\parallel}, \omega+i \eta)\right]/A({\bold k}_{\parallel}, \omega)
\end{eqnarray}
where ${\boldsymbol \sigma}$ are the Pauli matrixes. 

\subsection{Algorithm for searching nodal points/lines}
Nodal point is a gapless point between the highest valence band and the lowest conduction band. In terms of  degeneracy character, nodal point could be classified into Weyl, Triple, Dirac, hyper-Dirac point with 2-fold degeneracy, 3-fold degeneracy, 4-fold degeneracy or higher degeneracy respectively. They also can be sorted into nodal point and nodal line by the connectivity between them.
Searching the Weyl/Dirac points and the nodal-line structures is very important for such nodal systems. Some symmetry protected nodal points or nodal lines which  are located in high symmetry lines or mirror planes are easy to find. While the other nodes which are located  anywhere in the BZ need more efforts to be found. Here we introduce an algorithm trying to find all the nodal points. 

Basically, node points are local minima of the energy gap function in 3D BZ. Local minima can be obtained by using some well known multidimensional minimization methods, e.g., {\it Nelder and Mead's Downhill Simplex Method}~\cite{Nelder01011965}, {\it Conjugate Gradient Methods}~\cite{hazewinkel1994encyclopaedia}, {\it Quasi-Newton Methods}~\cite{press2007numerical} et al.. However, the local minimum obtained from those methods depends on a initial point. One initial point  gives only one local minimum. So, in order to find all the nodes, we have to choose different initial points in the whole 3D BZ. \texttt{WannierTools} takes a uniform mesh of the 1st BZ as a set of initial points for the {\it Nelder and Mead's Downhill Simplex Method}~\cite{Nelder01011965}. Eventually, the nodes will be selected out from a set of local minima. It is easy to check the convergence of the number of nodes by increasing the initial point mesh. This algorithm is very suitable for high throughput search of new Weyl, Dirac semimetals and nodal-line metals.  It has been checked to be very efficient to find Weyl points in WTe2~\cite{Soluyanov2015,Bruno2016}, MoTe2~\cite{Tamai2016}.  

\section{Capabilities of \texttt{WannierTools}} \label{sec:functions}
There are two kinds of tasks that \texttt{WannierTools} can do to  study novel topological materials. a. One is to get the topology of materials' band structure. b. The other one is to explore the properties of surface states corresponding to the bulk topology. For part a, we need to study the bulk band structure, 3D Fermi surface, density of state (DOS) to check whether the bulk material is a band insulator or a metal. Further, WCCs calculations are applied  to get the  $\mathbb{Z}_2$ topological index or Chern number  for band insulators, the nodes searching algorithm and the energy gap function calculation are applied to search for Weyl/Dirac point positions or nodal-line structures; The Berry phase and Berry curvature calculations are also aided to the classification of topology. After the topological classification is done, one can turn to part b, which means to study the bulk topology related properties, such as joint density of state (JDOS), which is related to the optical conductivity~\cite{book:Giuseppe}, electronic structure of the slab and wire systems, spin-texture of the surface states, Quasi-particle interference (QPI) pattern of surface states et al.. These two main capabilities of \texttt{WannierTools} are listed in Table \ref{tab:cap1} and  Table \ref{tab:cap2}. The meaning of control flags in Table \ref{tab:cap1} and  Table \ref{tab:cap2} is illustrated in the documentation, which is distributed with \texttt{WannierTools}.

\begin{center}
\begin{threeparttable}[htp]
\caption{Main capabilities of \texttt{WannierTools}: Bulk topology studies} \label{tab:cap1}
\begin{tabular}{l|l}
\hline
Control flag in \texttt{wt.in}   & Description\\
\hline 
BulkBand\_calc         & Energy bands for a  3D bulk system.    \\
BulkFS\_calc           & 3D Fermi surface in 1st BZ.    \\
FindNodes\_calc        & Locate Weyl, Dirac point positions and nodal line structures in 1st BZ.    \\
BulkGap\_plane\_calc   & Gap function in a 3D k plane.    \\
Wanniercenter\_calc    & WCCs~\cite{Soluyanov2011_z2} for a 3D k plane.    \\
BerryPhase\_calc       & Berry phase for a closed path in 3D k space.    \\
BerryCurvature\_calc   & Berry curvature in a 3D k plane.    \\
\hline\hline
\end{tabular}
\end{threeparttable}
\end{center}

\begin{center}
\begin{threeparttable}[htp]
\caption{Main capabilities of \texttt{WannierTools}: related responses from the bulk topology} \label{tab:cap2}
\begin{tabular}{l|l}
\hline
Control flag in \texttt{wt.in}   & Description\\
\hline 
Dos\_calc             & Density of state of a 3D bulk system \\
JDos\_calc          & Joint density of state~\cite{book:Giuseppe} of a 3D bulk system \\
SlabBand\_calc       & Energy bands of a 2D slab system.     \\
WireBand\_calc       & Energy bands of a 1D ribbon system.    \\
SlabSS\_calc          & Surface state spectrum along some k lines at different energies.    \\
SlabArc\_calc         & Surface state spectrum in the 2D BZ at a fixed energy.    \\
SlabQPI\_calc         & Quasi-particle interference (QPI)~\cite{Inoue1184} pattern of surface state.     \\
SlabSpintexture\_calc & Spin texture~\cite{PhysRevLett.111.066801} of surface state.    \\
\hline\hline
\end{tabular}\end{threeparttable}
\end{center}


\section{Installation and usage} \label{sec:install}
In this section, we will show  how to install and use the \texttt{WannierTools} software package.
\subsection{Get \texttt{WannierTools}}
 \texttt{WannierTools} is an open source free software package. It is released on Github under the GNU General Public Licence 3.0 (GPL), and it can be downloaded directly from the public code repository:\\
https://github.com/quanshengwu/wannier\_tools.
\subsection{Build \texttt{WannierTools}}
To build and install \texttt{WannierTools}, a Fortran 90 compiler, 
BLAS, and LAPACK linear algebra libraries are needed. An MPI-enabled Fortran 90 compiler is needed if you want to compile a parallel version. \texttt{WannierTools} can be successfully 
compiled using the state-of-art Intel Fortran compiler. Most of the MPI implementations,
such as MPICH, MVAPICH and Intel MPI are compatible with \texttt{WannierTools}. The downloaded \texttt{WannierTools} software package is likely a compressed file with a {\texttt{zip}} or {\texttt{tar.gz}} suffix. One should uncompress it firstly, then  move into the \texttt{wannier\_tools/soc} folder and edit the \texttt{Makefile} file to configure the compiling environment. It is noteworthy that one should set up the Fortran compiler, BLAS
and LAPACK libraries manually by modifying the following lines in \texttt{Makefile} file according to the user's particular system.
\begin{verbatim}
  f90 =  
  libs = 
\end{verbatim}
Once the compiling environment is configured, the executable binary \texttt{wt.x} will be compiled by typing the following command in the current directory (\texttt{wannier\_tools/soc}) 
\begin{verbatim}
   $ make
\end{verbatim}

\subsection{Running \texttt{WannierTools}}\label{sec:running}
Before running  \texttt{WannierTools}, the user must provide two files  {\texttt{wannier90\_hr.dat}}\footnote{The name of this TB file will be shown in the {\texttt{wt.in}} file. So it could be any name you like.} and {\texttt{wt.in}}.
The file called {\texttt{wannier90\_hr.dat}} containing the TB parameters has fixed format which is defined in software Wannier90~\cite{Wannier90cpc}. It can be generated by the software Wannier90~\cite{Wannier90cpc}, or generated by users with a toy TB model, or generated from a discretization of $k\cdot p$ model onto a cubic lattice. The other file {\texttt{wt.in}} is the master input file for  \texttt{WannierTools}. It is designed to be simple and user friendly. The details of {\texttt{wt.in}}  are described comprehensively in the documentation that contained within the {\texttt{WannierTools} distribution. An example file is provided in  \ref{appendix:input}. 

After putting  {\texttt{wt.in}} and {\texttt{wannier90\_hr.dat}} in the same folder, one can run it in single processor in the same folder like this
 \begin{verbatim}
   $ wt.x &
\end{verbatim}
or in multiprocessor 
\begin{verbatim}
   $ mpirun -np 4 wt.x &
\end{verbatim}

Some important information during the running process are written in {\texttt{WT.out}}, from which, you can check the running status. After the whole program is done, you would obtain two kinds of files other than {\texttt{WT.out}}. One is the data file suffixed with {\texttt{dat}}. and the other one is a plotting script for software {\it gnuplot} suffixed with {\texttt{gnu}}.  You can get nice plots with {\it gnuplot}~\cite{gnuplot}. Taking a bulk band structure calculation from the examples,  two files {\texttt{bulkek.dat}} and {\texttt{bulkek.gnu}} are accomplished after a successful running of \texttt{WannierTools}. A band structure plot \texttt{bulkek.png} will be generated with the following command    
\begin{verbatim}
gnuplot bulkek.gnu
\end{verbatim}

\section{Examples}\label{sec:examples}
In the past few years, {\texttt{WannierTools}} has been successfully applied in many projects, such as finding type-II Weyl semimetals  WTe$_2$~\cite{Soluyanov2015, Bruno2016}, MoTe$_2$~\cite{Tamai2016}, triple point metals~\cite{Zhu2016} ZrTe, TaN, nodal chain metals IrF4~\cite{Bzdusek2016}, topological phase in InAs$_{1-x}$Sb$_x$ ~\cite{Winkler2016} et al.. Besides, more and more groups notice this package, and becoming users. There are several examples in the {\texttt{wannier\_tools/examples}} directory like Bi$_2$Se$_3$, WTe$_2$, IrF$_4$.    MLWF TB Hamiltonians for those materials and the necessary input files for generating those Hamiltonians can be downloaded from the Github repository~\cite{wtgithub}. The detailed hands-on tutorials for those examples are listed in the wiki of Github ~\cite{wtwiki}.  In this paper,  a new series of topological materials called ternary silicides and ternary germanides TiPtSi, ZrPtSi, ZrPtGe, HfPtSi and HfPtGe ~\cite{C6DT00861E,Yashiro200051} are exhibited as  an example to show  how to study  topological properties of new materials with {\texttt{WannierTools}}. 

\subsection{Crystal structure and Band structure}
Ternary silicides and ternary germanides are crystalized with the orthorhombic TiNiSi type structure in a nonsymmorphic  orthorhombic space group No.62 (Pnma)~\cite{Nagata1998112,C6DT00861E}, containing three glide reflections $G_x=\{m_x|\frac{1}{2},\frac{1}{2},\frac{1}{2}\}$, $G_z=\{m_z|\frac{1}{2},0,\frac{1}{2}\}$, $\tilde{G}_y=\{m_y|0,\frac{1}{2},0 \}$ ~\cite{glidey},  three screw rotations $S_x=\{R_{2x}|\frac{1}{2},\frac{1}{2},\frac{1}{2}\}$, $S_z=\{R_{2z}|\frac{1}{2},0,\frac{1}{2}\}$, $S_y=\{R_{2y}|0,\frac{1}{2},0 \}$ and an inversion symmetry $I$. These materials seem to be interesting systems in the search for new superconducting intermetallic compounds~\cite{Yashiro200051}. Due to the same crystal structure and similar chemical properties, these materials show very similar band structures. In this paper, we are only focusing on  HfPtGe compound.

As  mentioned in section \ref{sec:running}, we need a TB model of HfPtGe for \texttt{WannierTools}. Firstly, a first-principle calculation within Vienna Ab initio Simulation Package (VASP) ~\cite{Kresse:1996,Kresse:1999} using Gamma centered K-points mesh  $8\times 13\times7$ and energy cut 360eV for plane wave expansions was performed. Then  the band structure and the partial density of states (PDOS)  shown in Fig \ref{fig:band}c and Fig \ref{fig:band}d were analyzed, where the PDOS indicate that the relevant orbitals close to the Fermi level are dominated by Hf 5d orbitals and Pt 5d orbitals, besides, they are also hybridized with Ge 4p  and Pt s orbitals. In the end, a 112-band MLWF TB Hamiltonian with Hf 5d, Pt 6s 5d and Ge 4p as projectors are constructed with Wannier90. Fig.\ref{fig:band}c shows that the band structures calculated from the MLWF TB model  are quite fitted to the first-principle calculated band structures.  After the successful construction of MLWF TB model, the master input file \texttt{wt.in} is needed,  which is attached in the \ref{appendix:input}.

\begin{figure}[!htp]
{\includegraphics[clip,height=3.2cm]{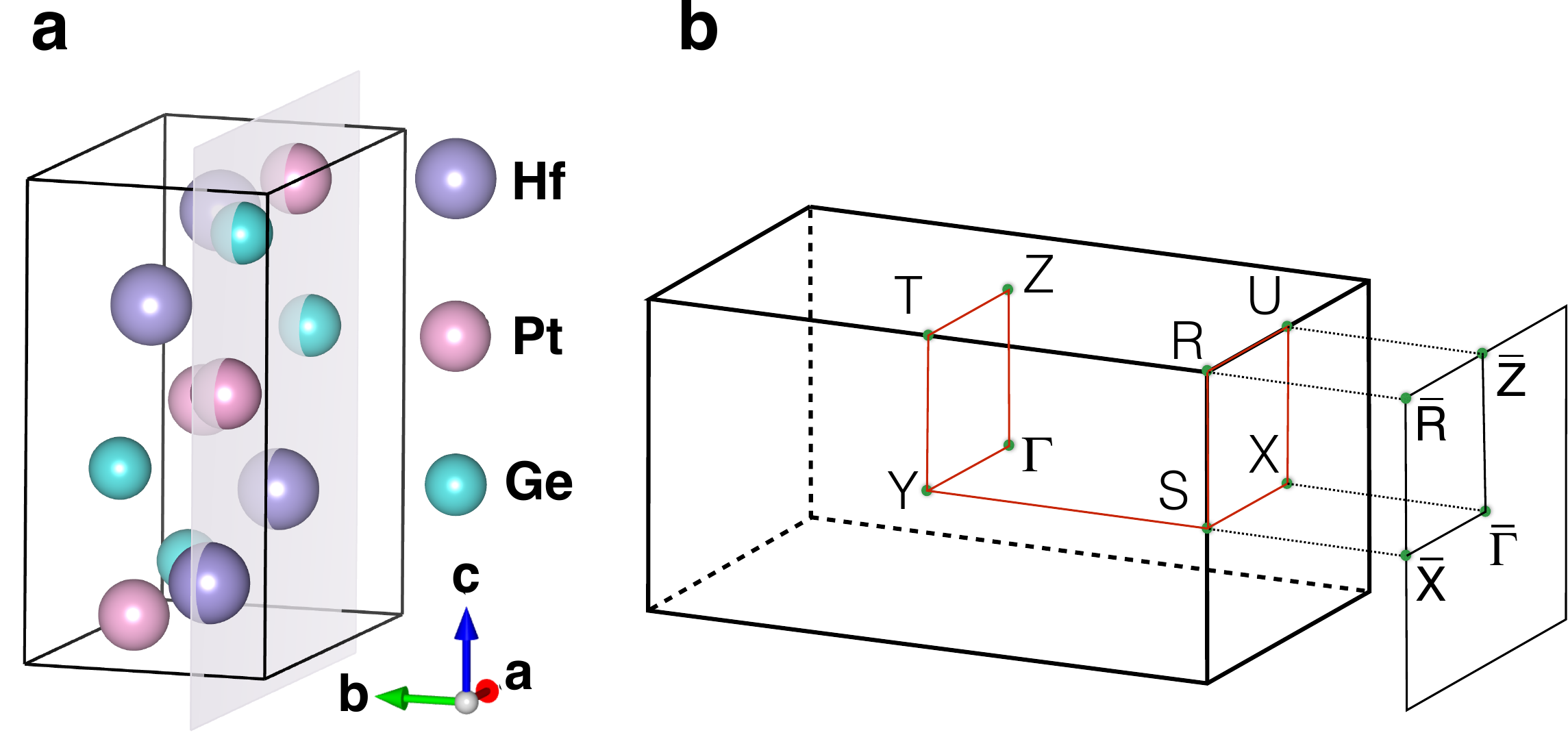}}\ \ 
{\includegraphics[clip,height=3.2cm]{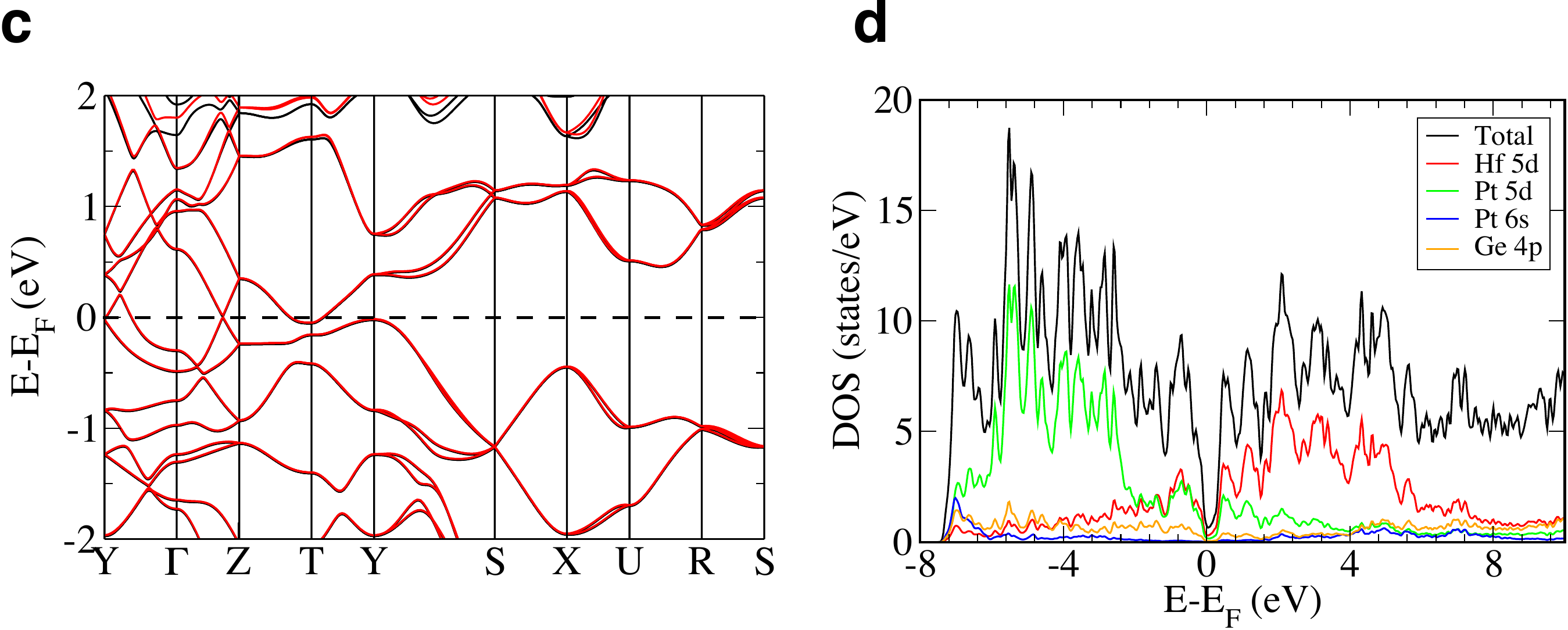}}
\caption{\label{fig:band}
a. Crystal structure of HfPtGe. The grey plane is (010) plane. b.  3D Brillouin Zone (BZ) and 2D BZ for (010) surface c. Band structure of HfPtGe with SOC, The red lines are from Wannier TB model, the black lines are from first-principle calculation. d.  Partial density of states (PDOS)}
\end{figure}

\subsection{Energy gap shape}
There are two "Dirac like" cones along Y-$\Gamma$-Z direction which are shown in the band structure plot Fig.~\ref{fig:band} (c). From the literature, we know that such cones could be originated from a nodal line structure without SOC~\cite{PhysRevLett.115.036807}. Calculations of the energy gap at $\text{k}_\text{y}=0$ plane  both for without SOC  and with SOC cases were performed to study the details of  positions of these "Dirac like" cones. The results are shown in Fig.\ref{fig:gap}a and  Fig.\ref{fig:gap}b, where there is a gapless nodal line protected by the $\tilde{G_y}$ mirror symmetry~\cite{PhysRevLett.115.036807} in the $\text{k}_\text{y}=0$ plane without SOC , while, the gapless line will be gapped when SOC turns on. One thing should mentioned is that the SOC strength of this material is very weak, and the opened gap is not very big. The smallest gap is about 0.1meV where the k points are located at ($\pm$0.4, 0.0, $\pm$0.229). 

We can further study the nodal line distribution in  momentum and energy space, which shown in Fig.\ref{fig:gap}c. It is clearly shown that the nodal line is not in the same energy plane. There are six nodes crossing the Fermi level. In such case, there are electron pockets and hole pockets which link together on the Fermi surface at the same time, which are shown in Fig \ref{fig:gap} (d). Those compensated hole and electron pockets will cause extremely large positive magnetoresistance~\cite{Ali_2014}.   

The related settings in the master input file \texttt{wt.in} for this section are as follows

\begin{verbatim}
    &CONTROL
    BulkGap_plane_calc = T
    BulkFS_calc        = T
    /
    
    &PARAMETERS
    NK1= 101
    NK2= 101
    NK3= 101
    /

    KPLANE_BULK
    -0.50  0.00 -0.50   ! Original point for 3D k plane
     1.00  0.00  0.00   ! The first vector to define 3d k space plane
     0.00  0.00  1.00   ! The second vector to define 3d k space plane
\end{verbatim}

\begin{figure}[!htp]
\begin{center}
{\includegraphics[clip ,height=3.5cm]{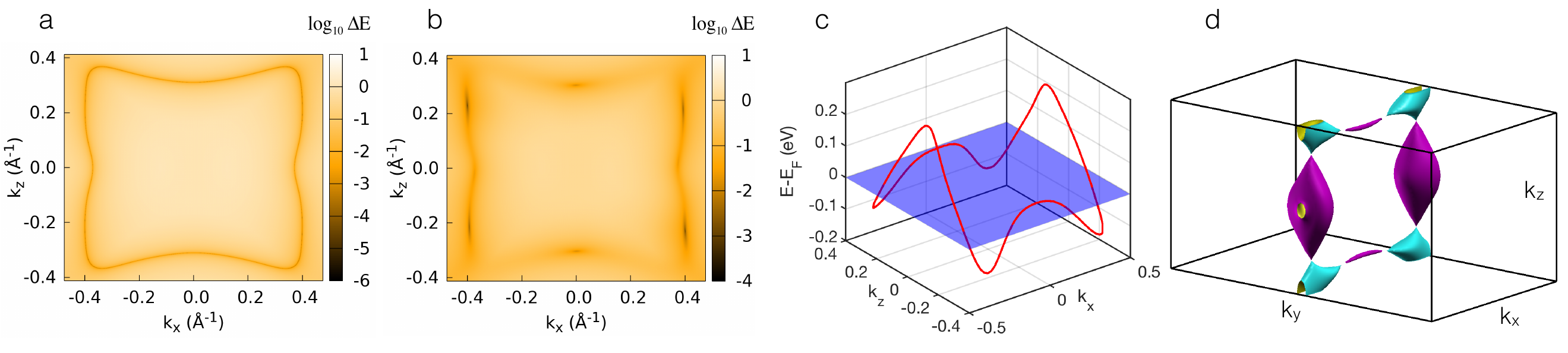}}
\caption{\label{fig:gap}
Energy gap $\Delta$E($\text{k}_\text{x}$, $\text{k}_\text{z}$) between the lowest conduction band and the topest valence band at $\text{k}_\text{y}=0$ plane. a. Without SOC, it is a nodal loop. b. With SOC, The nodal loop will be gapped out.   c. The nodal loop distribution in the momental space $\text{k}_\text{x}$-$\text{k}_\text{z}$  and energy space E. The blue plane is a energy fixed plane E={$\text{E}_\text{F}$} . d. Fermi surface plot, purple and cyan pockets represent hole and electron pockets respectively.  }
\end{center}
\end{figure} 

\subsection{Wannier charge center}
From the gap shape calculation for the whole BZ in the SOC case, we can conclude that HfPtGe is a semimetal with  a continuous finite energy gap between electron-like and hole-like bands. Similar to  classification of band insulators~\cite{Fu2007Z2}, $\mathbb{Z}_2$  topological indices $(\nu_0,\nu_1\nu_2\nu_3)$  are still appropriate for such a semimetal. $\mathbb{Z}_2$ number of a bulk material can be obtained through calculations of WCCs in six time reversal invariant planes $\text{k}_\text{x}=0, \pi$, $\text{k}_\text{y}=0, \pi$ and $\text{k}_\text{z}=0, \pi$ plane. The results calculated by \texttt{WannierTools} are shown in Fig.~\ref{fig:wcc}. It shows that the $Z_2$ invariant numbers are 1 for $\text{k}_\text{x}=0, \text{k}_\text{y}=0, \text{k}_\text{z}=0$ plane, while  zeros for other planes. Eventually, the topological index is $(1, 000)$, which indicates that HfPtGe is a "strong" topological material in all three reciprocal lattice directions. The related settings for this section in \texttt{wt.in} are as follows.
\begin{verbatim}
    &CONTROL
    Z2_3D_calc = T
    /
\end{verbatim} 

\begin{figure}[!htp]
\begin{center}
{\includegraphics[clip,height=6cm]{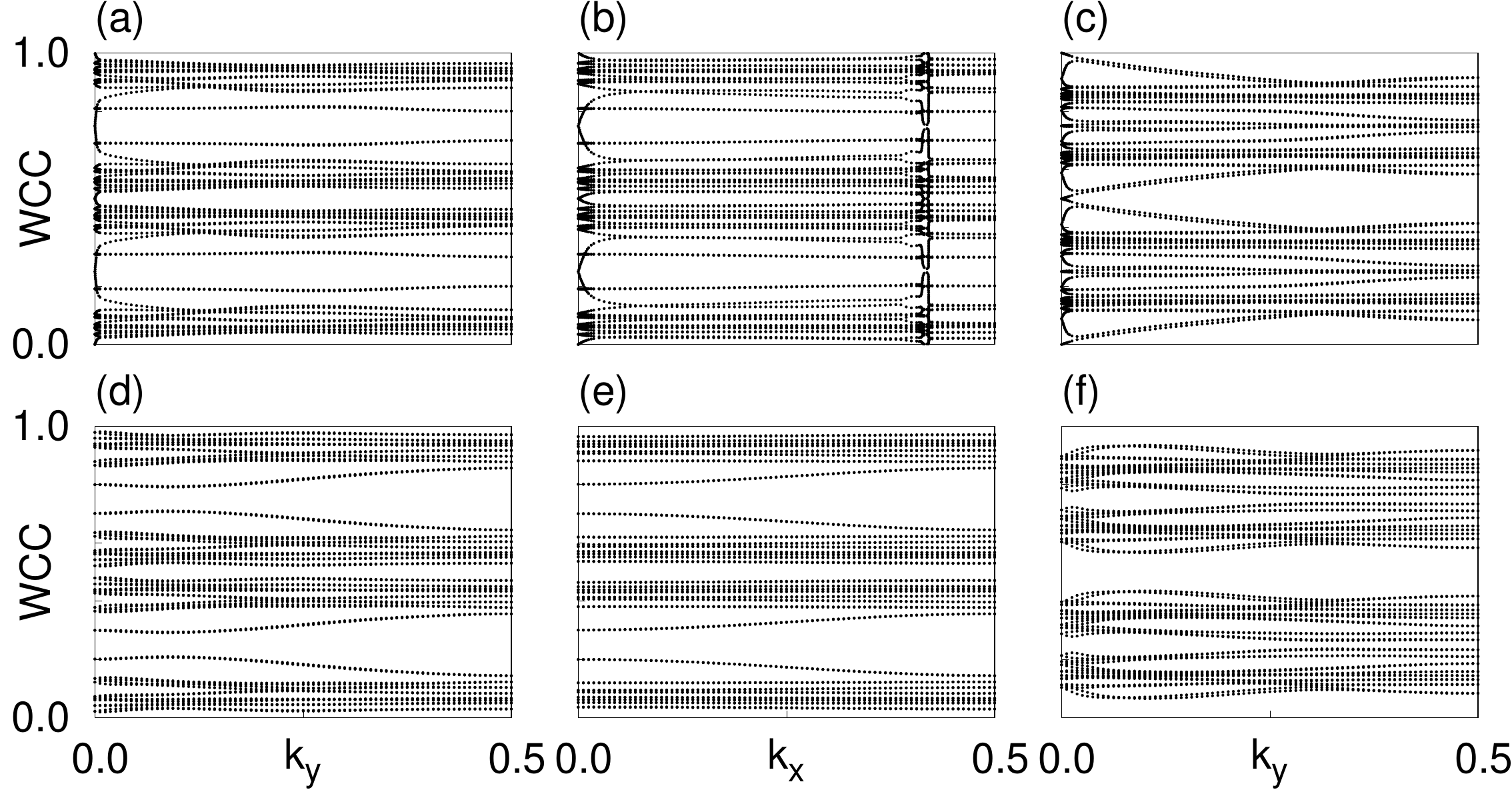}}
\caption{\label{fig:wcc}
a. Wannier charge center evolution for the time-reversal invariant planes of HfPtGe. (a)  $\mathbb{Z}_2=1$ for  $k_x$ = 0 (b)  $\mathbb{Z}_2=1$ for $k_y$ = 0 (c)  $\mathbb{Z}_2=1$ for ${k_z}$ = 0 (d)  $\mathbb{Z}_2=0$ for $k_x$ =$\frac{\pi}{a}$ (e)  $\mathbb{Z}_2=0$ for $k_y = \frac{\pi}{b}$  (f)  $\mathbb{Z}_2=0$ for  $k_z = \frac{\pi}{c}$}. 
\end{center}
\end{figure}

\subsection{Surface state spectrums}
Due to the bulk-edge correspondence, there should be topologically protected surface states of any cuts of surface for a strong topological material~\cite{Fu2007Z2}. Here we study the (010) surface which is shown as a grey plane in Fig.\ref{fig:band}a. The surface state spectrums calculated by \texttt{WannierTools} are shown in Fig.~\ref{fig:ss}a,\ref{fig:ss}b. For a 3D strong topological insulator, there is surface Dirac cone at the $\Gamma$ point~\cite{Zhang2009}. Indeed, there is a Dirac like cone of HfPtGe at the $\Gamma$ point. However, the dispersion of this cone is highly anisotropic and even tiled in the momentum space. Such tiled cone is result from  that the nodal line is located at different energies. Fig. \ref{fig:ss}b shows a $\text{E}=\text{E}_\text{F}$ iso-energy plot  of the surface state spectrum. From this plot, we can learn that the surface states originate from the k points that have the smallest gap. The related settings for this section in \texttt{wt.in} are as follows.
\begin{verbatim}
    &CONTROL
    SlabSS_calc  = T   
    SlabArc_calc = T
    /
    
    &PARAMETERS
    E_arc = 0.0    ! Fixed energy Fermi arc calculation
    /
    
    KPATH_SLAB
    4        ! numker of k lines for a slab system
    X -0.50  0.00  G  0.00 0.00  ! k path for a slab system
    G  0.00  0.00  Z  0.00 0.50
    Z  0.00  0.50  R -0.50 0.50
    R -0.50  0.50  X -0.50 0.00
    
    KPLANE_SLAB
    -0.5 -0.5
     1.0  0.0
     0.0  1.0
\end{verbatim} 

\begin{figure}[!htp]
\begin{center}
{\includegraphics[clip,height=4.5cm]{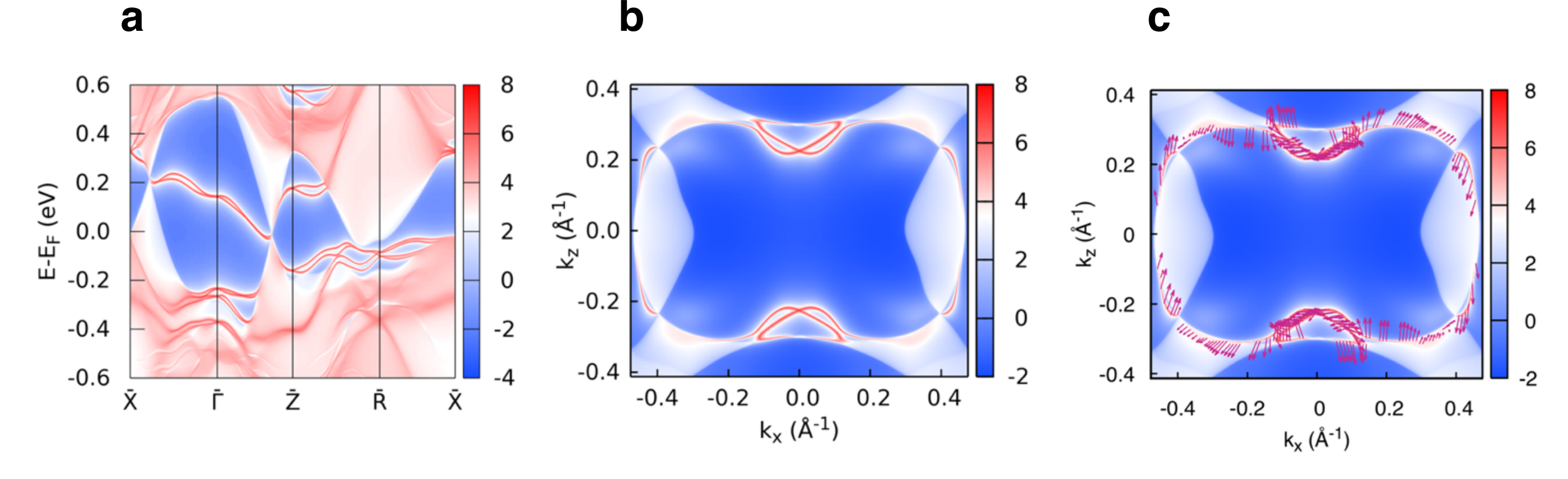}}
\caption{\label{fig:ss}
Surface state spectrum (SSS) of HfPtGe in consideration of SOC. a. SSS along high symmetry k-line at different energies. b. SSS in 2D BZ at a fixed energy $E-E_F=0$. c. Spin texture for the SSS in (b).}
\end{center}
\end{figure} 

\subsection{Surface state spin texture}
The key ingredient to generate topological non-trivial properties is the SOC interaction, which have a similar behavior as a Lorentz force in quantum Hall states.  The SOC interaction will make a spin and a momentum locked to each other, forming a spin texture in momentum space. For different topological phase, the spin texture will be different~\cite{Zhang2013}. In topological insulator Bi$_2$Se$_3$, the spin texture is Dirac type. The spin texture calculated by \texttt{WannierTools} for HfPtGe is shown in Fig.\ref{fig:ss}c. The related settings for spin texture calculations in \texttt{wt.in} are as follows.
\begin{verbatim}
    &CONTROL
    SlabSpintexture_calc = T
    /
    
    KPLANE_SLAB
    -0.50 -0.50
     1.00  0.00 
     0.00  1.00
\end{verbatim} 

\section{Conclusions}
In conclusion, we presented  an open-source software package called \texttt{WannierTools}. It is very user-friendly and is written in Fortran90, using MPI techniques to get excellent performance in computer cluster. We showed  how to use this software package to identify the topological properties for a new material and to get the surface state spectrum which can be compared with experimental data. As an example, we explored a new topological  material HfPtGe, which was identified as a Dirac nodal line semimetal.
 
\section*{Acknowledgments}
QSW would like to thank Xi Dai, Zhong Fang, Lei Wang, Li Huang, Dominik Gresch, Zhida Song for helpful discussions. Especially, QSW appreciates Rui Yu and Haijun Zhang for their kindly helps at the beginning of this project.   QSW, AAS, MT were supported by Microsoft Research, and the Swiss National Science Foundation through the National Competence Centers in Research MARVEL and QSIT, ERC Advanced Grant SIMCOFE. QSW was also supported by the National Natural Science Foundation of China (11404024). SNZ was supported by NSF-China under Grants No. 11074174. HFS was supported by National High Technology Research and Development Program of China under Grant 2015AA01A304, and Science Challenge Project No. JCKY2016212A502. This job was started in IOP CAS, finished in ETH Zurich.

Note added.—Recently, Ref. ~\cite{Singh2017} appeared, discussing topological properties of the same class of HfGePt.

\appendix

\section{wt.in for HfPtGe}\label{appendix:input}
\begin{verbatim}
&TB_FILE
Hrfile = "wannier90_hr.dat"
/

&CONTROL
BulkBand_calc         = T
BulkFS_calc           = T
BulkGap_plane_calc    = T
Z2_3D_calc            = T
SlabSS_calc           = T
SlabArc_calc          = T
SlabSpintexture_calc  = T
/

&SYSTEM
NumOccupied = 64
SOC = 1
E_FERMI = 8.4551
/

&PARAMETERS
E_arc = 0.0
OmegaNum = 100
OmegaMin = -0.6
OmegaMax =  0.6
Nk1 = 101
Nk2 = 201
Nk3 = 101
NP = 2
Gap_threshold = 0.05
/

LATTICE
Angstrom
6.6030000   0.0000000   0.0000000
0.0000000   3.9500000   0.0000000
0.0000000   0.0000000   7.6170000

ATOM_POSITIONS
12
Direct
  Hf  0.029900   0.250000   0.186300
  Hf  0.470100  -0.250000   0.686300
  Hf -0.029900   0.750000  -0.186300
  Hf  0.529900   0.250000   0.313700
  Ge  0.750600   0.250000   0.621500
  Ge -0.250600  -0.250000   1.121500
  Ge -0.750600   0.750000  -0.621500
  Ge  1.250600   0.250000  -0.121500
  Pt  0.142500   0.250000   0.561700
  Pt  0.357500  -0.250000   1.061700
  Pt -0.142500   0.750000  -0.561700
  Pt  0.642500   0.250000  -0.061700

PROJECTORS
 5 5 5 5 3 3 3 3 6 6 6 6
Hf dxy dyz dxz dx2-y2 dz2
Hf dxy dyz dxz dx2-y2 dz2
Hf dxy dyz dxz dx2-y2 dz2
Hf dxy dyz dxz dx2-y2 dz2
Ge px py pz
Ge px py pz
Ge px py pz
Ge px py pz
Pt s dxy dyz dxz dx2-y2 dz2
Pt s dxy dyz dxz dx2-y2 dz2
Pt s dxy dyz dxz dx2-y2 dz2
Pt s dxy dyz dxz dx2-y2 dz2

SURFACE
 1  0  0
 0  0  1
 0 -1  0

KPATH_BULK
 9
Y -0.50000  0.00000  0.00000     G  0.00000  0.00000  0.00000
G  0.00000  0.00000  0.00000     Z  0.00000  0.00000  0.50000
Z  0.00000  0.00000  0.50000     T -0.50000  0.00000  0.50000
T -0.50000  0.00000  0.50000     Y -0.50000  0.00000  0.00000
Y -0.50000  0.00000  0.00000     S -0.50000  0.50000  0.00000
S -0.50000  0.50000  0.00000     X  0.00000  0.50000  0.00000
X  0.00000  0.50000  0.00000     U  0.00000  0.50000  0.50000
U  0.00000  0.50000  0.50000     R -0.50000  0.50000  0.50000
R -0.50000  0.50000  0.50000     S -0.50000  0.50000  0.00000

KPATH_SLAB
4
X -0.50 0.00 G  0.00 0.00
G  0.00 0.00 Z  0.00 0.50
Z  0.00 0.50 R -0.50 0.50
R -0.50 0.50 X -0.50 0.00

KPLANE_SLAB
-0.5 -0.5
 1.0  0.0
 0.0  1.0

KPLANE_BULK
-0.50  0.00 -0.50
 1.00  0.00  0.00
 0.00  0.00  1.00

KCUBE_BULK
-0.50 -0.50 -0.50
 1.00  0.00  0.00
 0.00  1.00  0.00
 0.00  0.00  1.00
\end{verbatim}

\section*{References}

\bibliography{refs}

\begin{thebibliography}{71}
\bibitem [{sur()}]{surface}%
  \BibitemOpen
  \href@noop {} {\bibinfo  {journal} {Note: For 3D system, it is surface. For
  2D system, surface represents the edge. For the simplicity, we use surface in
  the whole article.}\ }\BibitemShut {NoStop}%
\bibitem [{gli()}]{glidey}%
  \BibitemOpen
  \href@noop {} {\enquote {\bibinfo {title} {The reason that i put a $\sim$
  onto $\tilde{G}_y$ is that the shifting vector in $\tilde{G}_y$ is
  perpendicular to the mirror plane. the square of $\tilde{G}_y$ is the same as
  the square of $m_y$. in this regard, we can call $\tilde{G}_y$ a mirror
  plane.}}\ }\BibitemShut {NoStop}%
\end{thebibliography}

\end{document}